\documentclass[journal]{IEEEtran}
\ifCLASSINFOpdf
\else
\fi
\begin{document}
%
\title{Introduction to the Issue on Deep Learning for Image/Video Restoration and Compression}
%
%

\markboth{IEEE JOURNAL OF SELECTED TOPICS IN SIGNAL PROCESSING,~Vol.~15, No.~2, FEBRUARY~2021}%
{Shell \MakeLowercase{\textit{et al.}}: Bare Demo of IEEEtran.cls for IEEE Journals}
%



\maketitle



%
\IEEEpeerreviewmaketitle

\section{Introduction}

\IEEEPARstart{T}{he} huge success of deep-learning--based approaches in computer vision has inspired research in learned solutions to classic image/video processing problems, such as denoising, deblurring, dehazing, deraining, super-resolution (SR), and compression. Hence, learning-based methods have emerged as a promising nonlinear signal-processing framework for image/video restoration and compression. 

Recent works have shown that learned models can achieve significant performance gains, especially in terms of perceptual quality measures, over traditional methods. Hence, the state of the art in image restoration and compression is getting redefined.        
This special issue covers the state of the art in learned image/video restoration and compression to promote further progress in innovative architectures and training methods for effective and efficient networks for image/video restoration and compression.

In the following, we provide a short overview of the state of the art in learned image and video processing Section~\ref{sota}.  Section~\ref{articles} introduces the articles in this issue. Finally, we provide the outlook for future directions in Section~\ref{outlook}.

\section{Overview of the State-of-the-Art}
\label{sota}

\subsection{Image/Video Restoration and Super-resolution}
Many researchers reported results that exceed the state~of the art in image/video restoration and SR by a wide margin via supervised learning using pairs of ground-truth (GT) images/video and degraded or low-resolution (LR) images/video generated by known degradation models, such as bicubic downsampling. However, there is need for further research and room for improvement in at least three key areas: generalization of these results to real-world problems, efficiency of the solutions, and perceptual optimization of the results.

Most existing image restoration/SR methods assume a pre-defined degradation process from a GT image/video to a degraded/LR one, which can hardly hold true in real imaging with complex degradation types. To fill this gap, growing attention has been paid in recent years to approaches for unknown degradations, namely real-world SR or blind SR. We~can roughly divide these methods into four groups: The~first group of methods utilize an external dataset to learn a SR model well adapted to a large set of downsampling kernels, such as IKC~\cite{jgu2019}, SRMD~\cite{kzha2018}, or USRNet~\cite{zhang2020}. Another group of methods leverage the internal statistics within a single image derived from the degradation model, thus requiring no external dataset for training, like ZSSR~\cite{shoc2018} and DGDMLSR~\cite{xche2020}. The third group resorts to implicit modeling, which defines the~degradation process implicitly through a data distribution~\cite{ignatov2018,anoosheh2019,lugmayr2019}. Particularly, these methods utilize data-distribution learning with Generative Adversarial Networks (GANs)~\cite{goodf2014} to grasp the implicit degradation model possessed within dataset, like WESPE~\cite{ignatov2018}, FSSR~\cite{frit2019} and CinCGAN~\cite{yyuan2018}. The last group directly builds real image datasets with input-output pairs for specific applications, such as DPED~\cite{ignatov2017}, RealSR~\cite{cai2019}, Zurich RAW-to-RGB~\cite{ignatov2020} and DRealSR~\cite{wei2020}. These new datasets make it possible to take advantage of existing supervised-learning methods in real-world applications.  

For real-world applications besides dealing with data captured in uncontrolled or challenging conditions, the restoration/SR solutions need to be run-time, memory and energy efficient and to run on constrained hardware~\cite{aibenchmark2018}. In a pioneering work, Ronneberger et al.~\cite{ronneberger2015} introduced U-Net, a widely adopted efficient neural design for image to image mapping. Since then tremendous progress has been achieved in Neural Architecture Search (NAS)~\cite{elsken2019}. However, while very efficient architectures have been optimized for tasks such as image classification (MobileNetV3~\cite{mobilenetv3}), solutions are sought for image restoration/SR tasks as shown in the AIM 2020 challenge on efficient SR~\cite{AIM2020}.

Another active area of research is perceptual image restoration and SR. Variations of the GAN architecture have been proposed for various low-level--vision tasks to obtain perceptually better results with more texture details. In a pioneering work, Ledig et al.~\cite{ledig2017} have proposed SRGAN model that could generate photo-realistic images in SR tasks. Ignatov et al.~\cite{ignatov2017,ignatov2018} proposed to use perceptual losses and GANs to learn from paired or unpaired images to enhance the images from a smartphone camera to a DSLR target camera quality. In~\cite{blau2018}, the authors made the observation that there is a trade-off between fidelity (measured by full-reference metrics) and perception (measured by no-reference metrics). In the PIRM 2018-SR Challenge~\cite{PIRM2018}, ESRGAN~\cite{wang2018} achieved state-of-the-art performance by improving the network architecture for the generator and loss functions. Benefiting from a learnable ranker, RankSRGAN~\cite{zhang2019} can optimize the generative network in the direction of any image quality assessment (IQA) metrics and achieves state-of-the-art performance. Although remarkable progress has been made, Gu et al.~\cite{gu2020} reveal that existing IQA method cannot objectively evaluate perceptual SR methods. In the newly-proposed IQA dataset, there is still a large gap between IQA methods and human labels.

\subsection{Image/Video Compression}

Much of the early work in applying learned models to compression  focused on image compression, starting with approaches that solely learned non-linear transformations of image inputs without learning corresponding probability models~\cite{tod2016}.  Subsequent, more effective approaches jointly learned  models of non-linear auto-encoders with models of the latent-variable probability distributions~\cite{balle2017,theis2017}. The model coupling was done by minimizing the Lagrangian formulation of the rate-distortion loss, using the learned probability model for the rate estimation and the decoded reconstruction from the scalar-quantized latent variables for the distortion.  Adding side information (``hyper priors'') to allow the probability models themselves to adapt locally resulted in a learned models that exceeded the performance of traditional image encoders (e.g., BPG)~\cite{balle2018}.  Extensions to the adaptive probability modeling include additional layers of side information about the probability models for the hyper-priors themselves, as well as autoregressive context models~\cite{min2018}.

Much of the previous work in learned video compression addressed the problem of replacing parts of standard compression systems (e.g. HEVC) with learned components~\cite{zhao2016,dai2017,li2018,wang2019,helle2019, gor2020,murn2020}. End-to-end optimized fully learned models have also shown promise. Some learned models based on uni-directional motion-compensation (low-latency)~\cite{ouy2019,rippel2019} have outperformed H.264 in PSNR and HEVC in MS-SSIM. The best performance to date in fully learned low-latency video compression uses a learned scale-space motion-compensation model~\cite{agus2020}. Recently, end-to-end optimized learned models based on bi-directional motion compensation have also shown competitive performance~\cite{reny2020,myil2020}.

\subsection{Point Cloud Denoising and Compression}
3-D point clouds (PC) are a collection of millions of points, where each point represents a specific 3-D coordinate of a scene, and associated color features. Raw 3-D PCs can be obtained by various acquisition devices or as output of 3-D reconstruction algorithms. Among many other applications, they are used as a scene representation for free view-point imaging and video. Compared to 3D meshes, they offer a simpler, denser, and closer-to-reality representation. However, raw 3-D PCs are typically contaminated with noise and outliers and the size of raw 3-D PC data is huge for storage and/or streaming. Hence, denoising and compression of 3-D PC are recent topics of significant interest.

Traditional methods for 3-D PC denoising typically rely on local surface fitting, local or non-local averaging, or on statistical modeling of the data and noise~\cite{han2017}. In contrast, deep learning offers a simple and universal data-driven approach for removing outliers and denoising 3-D PCs, corrupted with potentially very high levels of structured noise. 

Among many existing traditional approaches to 3-D PC compression, the MPEG-3DG (3D Graphics group) has standardized two different frameworks: i) Video-based Point Cloud Compression (V-PCC), and ii) Geometry-based Point Cloud Compression (G-PCC)~\cite{graz2020}. V-PCC considers compression of 2-D projections of 3-D PC to leverage the existing and future video compression technologies, as well as the established video eco-system. The reference model encoder achieves compression rates of 125:1 with good perceptual quality.
G-PCC considers 3-D geometry-driven approaches to provide efficient lossless and lossy compression. Recently, deep-learning based data-driven methods have started to achieve state of the art 3-D PC compression performance.

\section{Overview of the Articles}
\label{articles}
This special issue consists of 20 papers on recent advances in deep learning for image/video restoration and compression: 13 papers on image/video restoration and super-resolution, 5~papers on image/video compression, and 2 papers on point cloud processing. We provide a short introduction to these papers in the following.

\subsection{Image/Video Restoration and Super-resolution}

The paper "Degradation aware approach to image restoration using knowledge distillation" is the first journal paper on application of knowledge distillation on image restoration. 
The authors present a new approach to handle image-specific and spatially-varying degradations that occur in practice, such as rain-streaks, haze, raindrops, and motion blur. They decompose the restoration task into two stages of degradation localization and degraded region-guided restoration, unlike existing methods that directly learn a mapping between the degraded and clean images.

In ``Color image restoration exploiting inter-channel correlation with a 3-stage CNN'' Cui et al. propose a 3-stage CNN for color image restoration tasks. In this framework, first the green component is  reconstructed, followed by the red and blue channels with parallel networks, then all the intermediate reconstructions are concatenated to generate the final result. This method is successful in three typical color image restoration tasks: color-image demosaicking, color compression artifact reduction, and real-world color image denoising. 

Another image restoration work ``A deep primal-dual proximal network for image restoration'' borrows idea from image classification tasks and proposes a primal-dual proximal network. Specifically, it reformulates a specific instance of the primal-dual hybrid gradient (PDHG) algorithm as a deep network with fixed layers. Each layer corresponds to one iteration of the primal-dual algorithm. Two learning strategies -- Full learning and Partial learning -- are also proposed for better optimization. The proposed DeepPDNet shows excellent performance on the several benchmark datasets for image restoration and super resolution.

The paper ``Semi-supervised landmark-guided restoration of atmospheric turbulent images'' considers restoration of atmospheric turbulent (AT) images. As there is no paired training dataset for AT images, especially with faces, this work proposes a semi-supervised method for jointly extracting facial landmarks and restoring degraded images. The proposed approach learns to generate AT images by combining the content from a clean image and turbulence information from AT images in an unpaired manner. It adopts heatmaps from the landmark localization network as an additional prior. Experiments demonstrate the effectiveness of the proposed network on both AT image restoration and landmark localization.

In the rain-removal task, Kui et al. (``Multi-level memory compensation network for rain removal via divide-and-conquer strategy'') leverage the divide and-conquer strategy by decomposing the learning task into several subproblems according to levels of texture richness. It produces a high-quality rain-free image by subtracting the predicted rain information from multiple subnetworks. Each subnetwork processes a specific sub-sampled image, sampled from the original rainy ones via the Gaussian kernel. Experiments show that the proposed MLMCN outperforms existing deraining methods on several benchmark datasets, and the high-level object detection task.

Also, Yasarla et al. (``Exploring Overcomplete Representations for Single Image Deraining using CNNs'') proposes a deraining solution called Over-and-Under Complete Deraining Network (OUCD). OUCD consists of two branches: one employing an overcomplete convolutional network architecture for learning local structures by restraining the receptive field of filters and another one employing U-net targeting global structures. The solution significantly improves over state-of-the-art on synthetic and real benchmarks.

Ning et al. (``Accurate and Lightweight Image Super-Resolution with Model-Guided Deep Unfolding Network'') propose an explainable approach toward SISR named model-guided deep unfolding network (MoG-DUN). 
MoG-DUN unfolds the iterative process of model-based SISR into a multi-stage concatenation of building blocks with three interconnected trainable modules (denoising, nonlocal-AR, and reconstruction). Experiments show improvements over existing model-based methods.

In ``Multi-scale image super-resolution via a single extendable deep network'' Zhang et al. propose a solution (MSWSR) addressing efficiency and arbitrary upscaling factors. MSWSR implements multi-scale SR simultaneously by
learning multi-level wavelet coefficients of the target image.
Structurally, MSWSR is composed of one CNN part for low frequencies and one extendable RNN part for high frequencies and multiscale SR. A side window kernel is proposed for efficiency.

In ``WDN: A Wide and Deep Network to Divide-and-Conquer Image Super-resolution'', Singh and Mittal propose to divide the SISR problem into multiple sub-problems and then solve/conquer them within a neural network design. Their introduced network architecture is much wider and is deeper than existing networks and employs a new technique to calibrate the intensities of feature map pixels. The advantages are demonstrated through extensive experiments.

In ``Multi-Grid Back-Projection Networks'' Navarrete Michelini et al. demonstrate the power of the Multi–Grid Back–Projection (MGBP) network architecture on image and video super-resolution tasks with fidelity and/or perceptual quality targets. MGBP combines a novel cross-scale residual block inspired by the iterative back–projection (IBP) algorithm and a multi-grid recursion strategy inspired by multi–grid PDE solvers to scale computational complexity efficiently with increasing output resolutions.

In ``LSTM-DNN Based Autoencoder Network for Nonlinear Hyperspectral Image Unmixing'', Zhao et al. address the problem of blind hyperspectral unmixing by proposing a non-symmetric autoencoder network to fully exploit the spectral and spatial correlation information. LSTM captures spectral correlation information, a spatial regularization improves the spatial continuity of results, while an attention mechanism further enhance the unmixing performance. The effectiveness of the proposed method is validated on synthetic and real data.

``Uncertainty-Aware Semantic Guidance and Evaluation for Image Inpainting'' address the problem of filling in missing irregularly shaped areas of an image, a problem that arises in practice when trying to recover an image that has an overlay (e.g., super-imposed text) or a foreground object that is being synthetically removed or when trying to create a different viewpoint of a scene (in newly dis-occluded areas).  The approach that is taken is to iteratively evaluate inpainted visual contents as well as a structural segmentation mask.  The approach surpasses other state-of-the-art approaches, in terms of clear boundaries and photo-realistic textures.

The paper ``Deep energy: Task driven training of deep neural networks" offers an unsupervised training approach using task-specific energy functions, where the proposed solution is better than the one obtained by a direct minimization of the energy function due to added regularization property of deep neural networks.

\subsection{Image/Video Compression}

Breakthroughs in modeling latent-variable probability distributions jointly with parameterized non-linear transformations [16, 17] were what was needed to allow learning-based approaches to image compression to quickly surpass the performance achieved by more traditional models.   ``Nonlinear transform coding'' is the first journal paper with comprehensive coverage of latent-variable RD-curve optimization for nonlinear-transform coding. 


The next two papers focus on different approaches to intra-frame block compression.  ``Intra-frame coding using a conditional autoencoder'' introduces an auto-encoder approach to mode-selection for predicting intra-frame image/video blocks.  The learned latent-space variable is itself the prediction-function index (replacing the mode-index used in classic intra-frame block coding) and the context pixels condition both parts of the auto-encoder architecture.  Cross-channel prediction is provided between the luma and chroma encodings, to avoid the need to separately send the latent-variables for the chroma channels.  The results improve the Bjøntegaard delta rate (BD-rate) for both luma and chroma channels compared to previous state-of-the-art.
The second of these papers, ``Attention-based neural networks for chroma intra prediction in video coding'', also looks at intra-frame chroma prediction but does so with a very different approach.  In this paper, a purely convolutional network is used with an attention layer~\cite{att2017} to cross-index between the (known) chroma boundary pixels and the (previously decoded) luma pixels within the block.  Since the network is purely convolutional, it is able to handle all block sizes (4x4, 8x8, and 16x16), reducing both the space required for the models and the average computation used across the video duration.

The next paper, ``MFRNet: A new CNN architecture for post-processing and in-loop filtering,'' also looks at leveraging convolutional neural-networks within the framework of a traditional video compressor, but this time for in-loop and post-processing filtering.  The paper introduces a new neural-network architecture that allows reuse of early-layer representations throughout the remaining layers of the network.  The results show significant PSNR gains for both in-loop and post-processing.

Finally, ``Learning for video compression with recurrent auto-encoder and recurrent probability model'' presents a fully learning-based approach to video compression that outperforms the default-speed setting for x265, using recurrent probability models for the latent variables of the recurrent auto-encoder network that is used to encode the motion-compensated video frames.

\subsection{Point Cloud Denoising and Compression}
Finally, there are two papers on deep learning for point cloud processing, one on denoising and one on compression.

The paper entitled ``Learning robust graph-convolutional representations for point cloud denoising'' proposes a deep learning method that can simultaneously denoise a point cloud and remove outliers in a single model. The core of the proposed method is a graph-convolutional neural network able to efficiently deal with the irregular domain and the permutation invariance problem typical of point clouds.

The paper entitled ``Adaptive deep learning-based point cloud geometry coding'' is the first journal paper on point cloud compression. It presents a novel deep learning-based solution for point cloud geometry coding that divides the point cloud into 3D blocks and selects the most suitable available deep learning coding model to code each block, thus maximizing the compression performance.

\section{Outlook}
\label{outlook}
There are many compelling future research challenges still remain to be addressed. These include: i) learned models contain millions of parameters, which makes real-time inference on common devices a challenge, ii) it is difficult to interpret learned models or to provide performance bounds on results, iii) it is important to provide perceptual loss functions, for training, that accurately reflect human preferences, iv)~the~performance of learned models trained on synthetically generated data drops sharply on real-world images/video, where the quantity and quality of training data is limited, and v) exploiting temporal correlations for efficient and effective video restoration and compression is challenging.

We hope that this special issue broadly summarizes the current state
of the art in learned methods for image/video restoration and compression, and
inspires researchers to work on numerous future directions calling for deeper investigation.


\ifCLASSOPTIONcaptionsoff
  \newpage
\fi



%

\vspace{30pt}
A. MURAT TEKALP, Lead Guest Editor \\
\indent Dept. of Electrical and Electronics Eng.\\
\indent Koç University \\
\indent Istanbul, Turkey

\vspace{12pt}
MICHELE COVELL, Guest Editor \\
\indent Google Research  \\
\indent Mountain View, CA, USA

\vspace{12pt}
RADU TIMOFTE, Guest Editor \\
\indent ETH Zurich  \\
\indent Zurich, Switzerland

\vspace{12pt}
CHAO DONG, Guest Editor \\ 
\indent Shenzhen Institute of Advanced Technology  \\
\indent Shenzhen, China

\end{document}